\documentclass[fdp,a4paper,fleqn]{w-art}
\usepackage{times,cite,w-thm}
\theoremstyle{plain}

\theoremstyle{definition}

\usepackage[]{graphicx}

\begin{document}

\DOIsuffix{theDOIsuffix}

\Volume{55}
\Month{01}
\Year{2007}

\pagespan{1}{}

\keywords{string theory, singularities, time dependence, plane gravitational waves.}
\subjclass[pacs]{04A25
\qquad\parbox[t][2.2\baselineskip][t]{100mm}{%
  \raggedright
  11.25.-w, 98.80.Qc\vfill}}%

\title[Free string evolution across plane wave singularities]{Free string evolution across plane wave singularities}

\author[B. Craps]{Ben Craps\inst{1}}

\author[F. De Roo]{Frederik De Roo\inst{1,2,} 
\footnote{Corresponding author\quad E-mail:~\textsf{fderoo@tena4.vub.ac.be},
            Phone: +32\,2\,629\,3464\,,
            Fax: +32\,2\,629\,2276\\Aspirant FWO}}
\address[\inst{1}]{Vrije Universiteit Brussel and International Solvay Institutes, Pleinlaan 2, 1050 Brussels, Belgium}
\address[\inst{2}]{Universiteit Gent, IR08, Sint-Pietersnieuwstraat 41, 9000 Ghent, Belgium}

\author[O. Evnin]{Oleg Evnin\inst{1}}

\begin{abstract}

In these proceedings, we summarize our studies of free string propagation in (near-)singular scale-invariant plane wave geometries. We analyze the singular limit of the evolution for the center-of-mass motion and all excited string modes. The requirement that the entire excitation energy of the string should be finite excludes consistent propagation across the singularity, in case no dimensionful scales are introduced at the singular locus (in an otherwise scale-invariant space-time).

\end{abstract}
\maketitle                   


\section{Introduction}

The propagation of quantum strings in time-dependent and singular backgrounds, necessary to understand the nature of cosmological space-time singularities in quantum gravity, is not well understood yet. String propagation in plane gravitational waves has been studied extensively (see \cite{HorowitzSteif,PRT} among other publications) because these space-times exhibit some highly special properties: the structure of the Riemann tensor implies that all higher curvature invariants vanish, and therefore these backgrounds have no $\alpha'$ corrections in perturbative string theories \cite{HorowitzSteif}. The existence of a lightlike covariantly constant Killing vector ensures there is no string or particle creation and permits an analytic treatment of the string $\sigma$ model for exact plane waves (the light-cone Hamiltonian is quadratic and analytically solvable).

We will investigate the transition of a free string across a (one parameter) class of resolved gravitational plane waves that develop a singularity at the origin of time when the resolution parameter $\epsilon$ is taken to zero. We shall concentrate on exact plane waves whose singular limit ($\epsilon\rightarrow 0$) is given by the following scale-invariant profile:
\begin{equation}
ds^2=-2dX^+dX^- - F(X^+) \sum_{i=1}^d (X^i)^2 (dX^+)^2 + \sum_{i=1}^d(dX^i)^2, \hspace{5mm} F(X^+)= \frac{\lambda}{(X^+)^2}.\label{genppwave}
\end{equation}
This class of backgrounds arises as a Penrose limit of a broad class \cite{plimit} of space-time singularities. Positive values of $\lambda$ correspond to Friedmann-like Big Bang singularities, and negative values of $\lambda$ correspond to an infinite-expansion rather than an infinite-contraction singularity (``Big Rip'').

Consistency conditions for the space-time background in which a perturbative string theory is formulated take the form of supergravity equations of motion together with an infinite tower of $\alpha'$-corrections. For non-singular plane waves, all the $\alpha'$-corrections vanish automatically. For singular space-times, the question of background consistency conditions {\it at} the singular point is subtle. One approach\footnote{For an alternative approach based on analytic continuation in the complex $X^+$-plane see \cite{PRT}. Background consistency conditions do not
enter the considerations within this framework.}, advocated in \cite{HorowitzSteif}, to formulating string theory in the background (\ref{genppwave}) is to resolve the singular plane wave profile into a non-singular function, perform the necessary computations and see if the result has a meaningful singular limit. Therefore we want to construct a resolved function $F(X^+,\epsilon)$ to replace $F(X^+)$ in (\ref{genppwave}) in such a way that
\begin{equation}
\lim\limits_{\epsilon\to 0}F(X^+,\epsilon)=\frac{\lambda}{(X^+)^2}
\label{limt}
\end{equation}
everywhere away from $X^+=0$. We remove some of the ambiguity associated with such resolutions by demanding that the original scaling symmetry of the background (\ref{genppwave}) is recovered when the resolution is removed. This will happen if the resolved profile $F(X^+,\epsilon)$ does not depend on any dimensionful parameters other than the resolution parameter $\epsilon$. In this case, on dimensional grounds,
\begin{equation}
F(X^+,\epsilon)=\frac{\lambda}{\epsilon^2}\Omega(X^+/\epsilon).
\label{scaleinvres}
\end{equation}
The limit (\ref{limt}) will be recovered, if, for large values of $\eta$,
\begin{equation}
\Omega(\eta)\to \frac{1}{\eta^2}+O\left(\frac1{\eta^b}\right),\hspace{10mm}\mathrm{for\;some\;} b>2.
\label{asrefprof}
\end{equation}

The structure of the paper is as follows: first we review the Hamiltonian formulation of free string dynamics in the background (\ref{genppwave}). Then we recapitulate the main results of \cite{EvninNguyen} for the evolution of the center-of-mass motion across this plane wave singularity. We extend this analysis to the evolution of excited string modes. We conclude by discussing stringent conditions arising if one demands the total mass of the string to remain finite after it crosses the singularity. A more detailed discussion of this work can be found in \cite{fspw}.


\section{Free strings in plane waves}

String worldsheet fermions are free in plane wave backgrounds and we shall therefore concentrate on the bosonic part of the string action. We choose light-cone gauge $X^+=\alpha'p^+ t$ and gauge-fix the metric such that the coupling to the dilaton disappears (see e.g. \cite{PRT}). We rescale and rename $\epsilon\rightarrow \epsilon \alpha' p^+$, set $\alpha'=1$, and Fourier transform the $X^i$ coordinates:
\begin{equation}
X^i(t,\sigma)=X^i_0(t)+ \sqrt{2}\sum_{n>0} \left(\mathrm{cos}\left(n\sigma\right) X^i_{n}(t)+\mathrm{sin}\left(n\sigma\right) \tilde{X}^i_{n}(t)\right),
\end{equation}
We obtain a set of harmonic oscillator Hamiltonians,
\begin{align}
H&= \sum_{n=0}^\infty \sum_{i=1}^d H_{ni},\label{setham}\\
H_{0i}&=\frac{(P_{0i})^2}{2}+ \frac{\lambda}{\epsilon^2}\Omega(t/\epsilon) \frac{(X_{0}^i)^2}{2},\\
H_{ni}&=\frac{(P_{ni})^2+(\tilde{P}_{ni})^2}{2}+ \left(n^2 + \frac{\lambda}{\epsilon^2}\Omega(t/\epsilon)\right) \frac{(X_{n}^i)^2+  (\tilde{X}_{n}^i)^2}{2}.\label{HOham}
\end{align}

The Hamiltonian (\ref{setham}) is quadratic and the related Schr\"odinger wave function $\phi(t;X_n^i)$ can be found using WKB techniques, which are exact for quadratic Hamiltonians. Up to normalization, a basis of solutions, labelled by the initial condition $X_n^i(t_1)=X_{1,n}^i$, can be written as \cite{EvninNguyen}
\begin{equation}
\phi(t;X_n^i) \sim \prod_{n}\prod_{i=1}^d\frac{1}{\sqrt{\mathcal{C}(t_1,t)}}\mathrm{exp}\left(-\frac{i}{2 \mathcal{C}} \left[(X_{1,n}^i)^2\partial_{t_1}\mathcal{C}-(X_n^i)^2\partial_{t}\mathcal{C}+2 X_{1,n}^i X_n^i\right]\right),\label{basisofsols}
\end{equation}
where $\mathcal{C}(t_1,t)$ (suppressing the index $n$) is a solution to the ``classical equation of motion'' for the time-dependent harmonic oscillator Hamiltonian (\ref{HOham}):
\begin{equation}
\partial^2_{t}{\mathcal{C}}(t_1,t)+\left(n^2+\frac{\lambda}{\epsilon^2}\Omega(t/\epsilon)\right)\mathcal{C}(t_1,t)=0,\label{ceomC}
\end{equation}
with specified initial conditions
\begin{equation}
\mathcal{C}(t_1,t)|_{t_1=t}=0,\hspace{5mm}\partial_{t} \mathcal{C}(t_1,t)|_{t_1=t}=1.\label{initC}
\end{equation}
A useful representation is given by
\begin{equation}
\mathcal{C}(t_1,t_2)=\frac{1}{W[f,h]}\left(f(t_1) h(t_2) - f(t_2) h(t_1)\right),\label{repCfh}
\end{equation}
where $f(t)$ and $h(t)$ are two independent solutions to the differential equation under consideration, and the Wronskian is given by $W[f,h]=f\dot{h}-h\dot{f}$. To derive the singular limit of the wavefunction (\ref{basisofsols}) it is sufficient to study the singular limit of (\ref{ceomC}-\ref{initC}).


\section{The singular limit for the center-of-mass motion}
\label{0mode}

For the $n=0$ mode, we obtain as the ``classical equation of motion''
\begin{equation}
\ddot{X}+\frac{\lambda}{\epsilon^2}\Omega(t/\epsilon)X=0.\label{ceomZM}
\end{equation}
From (\ref{initC}) it follows that we need to study the $\epsilon\rightarrow 0$ limit of the solution that obeys the initial conditions
\begin{equation}
X(t_1)=0,\hspace{10mm}\dot{X}(t_1)=1,\hspace{10mm}t_1<0. 
\label{initcondX}
\end{equation}
The singular limit has been rigorously considered in \cite{EvninNguyen}. Essentially, a scale transformation $t=\eta\epsilon$ and $Y(\eta)=X(t)$ permits to remove all $\epsilon$-dependence from equation (\ref{ceomZM}) leading to:
\begin{equation}
\frac{\partial^2}{\partial \eta^2}Y(\eta)+{\lambda}\Omega(\eta)Y(\eta)=0.
\end{equation}
In the singular limit, finiteness for $X(t)$ evolving from $t_1<0$ to $t_2>0$ translates into finiteness for $Y(\eta)$ evolving from $\eta_1=t_1/\epsilon$ to $\eta_2=t_2/\epsilon$. 

For the specific asymptotics (\ref{asrefprof}) of our resolved profile (\ref{scaleinvres}), it can be shown \cite{EvninNguyen} that, in the infinite past and infinite future, the solutions approach a linear combination of two powers (denoted below $a$ and $1-a$, with $a$ being a function of $\lambda$). The condition for the existence of the singular limit can be deduced by considering two bases of solutions for (\ref{ceomZM}), one asymptotically approaching the two powers (dominant and subdominant) at $\eta\rightarrow -\infty$ (i.e. for all $t<0$ when $\epsilon\rightarrow 0$),
\begin{equation}
Y_{1-}(\eta)=|\eta|^{a} + o(|\eta|^{a}),\hspace{10mm}Y_{2-}(\eta)=|\eta|^{1-a} + o(|\eta|^{1-a}),
\label{Y-}
\end{equation}
and another one behaving similarly at $\eta\rightarrow +\infty$ (i.e. for all $t>0$ when $\epsilon\rightarrow 0$)
\begin{equation}
Y_{1+}(\eta)=|\eta|^{a} + o(|\eta|^{a}),\hspace{10mm}Y_{2+}(\eta)=|\eta|^{1-a} + o(|\eta|^{1-a}),
\label{Y+}
\end{equation}
with $a=1/2+\sqrt{1-4\lambda}/2$. The two bases are related by a $2\times 2$ matrix $Q(\lambda)$:
\begin{equation}
\begin{bmatrix}Y_{1-}(\eta)\\Y_{2-}(\eta)\end{bmatrix}=Q(\lambda)\begin{bmatrix}Y_{1+}(\eta)\\Y_{2+}(\eta)\end{bmatrix}.
\label{Q}
\end{equation}

The condition for the existence of a singular limit can be now deduced \cite{EvninNguyen}. A heuristic version of the argument goes as follows: we consider a solution $\tilde Y(\eta)$ with initial conditions given at $\eta_1=t_1/\epsilon$, that evolves towards $\eta_2=t_2/\epsilon$, and decompose it with respect to the basis \{$Y_{1-}$,$Y_{2-}$\} as $\tilde Y=C_1 Y_{1-}+C_2 Y_{2-}$. At $\eta_1$ we can use the asymptotic expansion (\ref{Y-}) and for generic initial conditions (including (\ref{initcondX})) at $\eta_1$ both terms should be of the same order, such that $C_1=O(\epsilon^a)$ and $C_2=O(\epsilon^{1-a})$. We use the relation (\ref{Q}) to write $\tilde Y$ in terms of the basis with asymptotics (\ref{Y+}). Combining the powers of $\epsilon$ we obtain
\begin{align}
\tilde Y(t_2/\epsilon)=\,Q_{21}(\lambda)t_2^{a}\,O(\epsilon^{1-2a})&\,+\,Q_{11}(\lambda)t_2^{a}\,O(\epsilon^0)\nonumber\\&\,+\,Q_{22}(\lambda)t_2^{1-a}\,O(\epsilon^0)\,+\,Q_{12}(\lambda)t_2^{1-a}\,O(\epsilon^{2a-1}).
\end{align}
Since $a>1/2$, this expression can only have an $\epsilon\to 0$ limit if $Q_{21}(\lambda)=0$. Generically, this will lead to a discrete spectrum for $\lambda$, the normalization of the overall plane wave profile (\ref{scaleinvres}).


\section{The singular limit for excited string modes}

The evolution of excited string modes is described by time-dependent harmonic oscillator equations
\begin{equation}
\frac{\partial^2}{\partial t^2} X(t) + \left(n^2+\frac{\lambda}{\epsilon^2}\Omega(t/\epsilon)\right) X(t)=0.\label{de}
\end{equation}
It again suffices to analyze the singular limit of $\mathcal{C}(t_1,t_2)$ defined by (\ref{ceomC}-\ref{initC}). Because $n^2$ is finite, it is natural to expect that it does not affect the existence of the singular limit, governed by the singularity emerging from $\Omega(t/\epsilon)$. This is indeed the case for positive $\lambda$, or $a<1$. For negative $\lambda$ unstable motion of the inverted harmonic oscillator
leads to divergences and the analysis becomes more subtle. Further details can be found in \cite{fspw}.

To derive $\mathcal{C}(t_1,t_2)$ for equation (\ref{de}) we use the following strategy: the differential equation (\ref{de}) is linear and at any $t=t_2$ the solution $X(t)$ can be written in terms of a ``transfer matrix'' $T$ that only depends on the initial and final times,
\begin{equation}
\begin{bmatrix} X(t_2)\\\dot X(t_2)\end{bmatrix}=T(t_1,t_2)\begin{bmatrix} X(t_1)\\\dot X(t_1)\end{bmatrix},\hspace{10mm}T(t_1,t_2)=\begin{bmatrix}-\partial_{t_i}\mathcal{C}(t_1,t_2)&\mathcal{C}(t_1,t_2)\\- \partial_{t_i}\partial_{t_f}\mathcal{C}(t_1,t_2)&\partial_{t_f} \mathcal{C}(t_1,t_2)\end{bmatrix}.
\end{equation}
The transfer matrix is completely determined once $\mathcal{C}(t_1,t_2)$ has been determined, and vice versa. We now divide the solution region into three sub-intervals, indicated in the following figure:

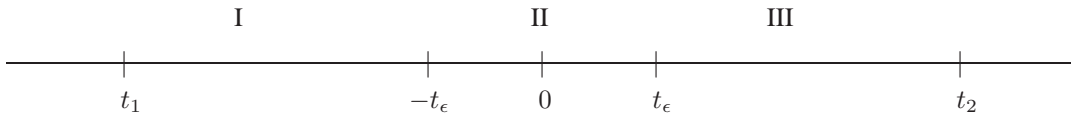
\setlength{\unitlength}{1mm}
\noindent\begin{picture}(140,20)
\put(0,10){\line(1,0){140}}
\put(30,15){\text{I}}
\put(69,15){\text{II}}
\put(100,15){\text{III}}
\put(15,9){\text{$|$}}
\put(55,9){\text{$|$}}
\put(85,9){\text{$|$}}
\put(125,9){\text{$|$}}
\put(15,4){\text{$t_1$}}
\put(53,4){\text{$-t_\epsilon$}}
\put(85,4){\text{$t_\epsilon$}}
\put(125,4){\text{$t_2$}}
\put(70,9){\text{$|$}}
\put(70,4){\text{$0$}}
\end{picture}

\noindent The boundaries of the near-singular region II depend on $t_\epsilon$ which is chosen to approach $0$ in the singular limit as 
\begin{equation}
t_\epsilon=\epsilon^{1-c}\tilde t^c,
\label{teps}
\end{equation}
with $\tilde t$ staying finite in relation to the ``moments of observation'' $t_1$ and $t_2$, and $c$ a number between 0 and 1. For each interval we now introduce a local transfer matrix $T_k$ for the evolution from the initial time of the interval to the final time of the interval. On each interval, we can write the transfer matrix $T_k$ in terms of a local $\mathcal{C}_k$. Using matrix multiplication to construct the full transfer matrix $T$ in terms of the $T_k$, we can deduce an expression for the $\mathcal{C}(t_1,t_2)$ of the complete interval, in terms of the local $\mathcal{C}_k$ of the three sub-intervals:
\begin{align}
&\mathcal{C}(t_1,t_2)=\mathcal{C}_{I}(t_1,-t_\epsilon)\partial_{t_i}\mathcal{C}_{II}(-t_\epsilon,t_\epsilon)\partial_{t_i}\mathcal{C}_{III}(t_\epsilon,t_2)-\partial_{t_f}\mathcal{C}_{I}(t_1,-t_\epsilon)\mathcal{C}_{II}(-t_\epsilon,t_\epsilon)\partial_{t_i}\mathcal{C}_{III}(t_\epsilon,t_2)\nonumber\\&-\mathcal{C}_{I}(t_1,-t_\epsilon)\partial_{t_i}\partial_{t_f}\mathcal{C}_{II}(-t_\epsilon,t_\epsilon)\mathcal{C}_{III}(t_\epsilon,t_2)+\partial_{t_f}\mathcal{C}_{I}(t_1,-t_\epsilon)\partial_{t_f}\mathcal{C}_{II}(-t_\epsilon,t_\epsilon)\mathcal{C}_{III}(t_\epsilon,t_2).
\label{Cexpr0}
\end{align}
With $\partial_{t_i}$ and $\partial_{t_f}$ we differentiate each $\mathcal{C}_k$ with respect to its first and second argument (evaluated at the initial and final time of the local interval).

We study the existence of the singular limit of $\mathcal{C}(t_1,t_2)$ as follows: for two linear differential equations related by a small perturbation we can establish a bound on the difference between perturbed and unperturbed solutions with the same initial conditions, using Gronwall's inequality (see e.g. \cite{gronwall}). This bound, of course, applies to each local $\mathcal{C}_k$. For each of the three sub-intervals introduced above, we consider a simplified differential equation that is a good approximation to equation (\ref{de}) on the corresponding interval:
\begin{itemize}
\item Region I and III: $\ddot{X}(t) + \left(n^2+ \lambda/t^2\right) X(t)=0$ (related to Bessel's equation);
\item Region II: $\ddot{X}(t) + \lambda/\epsilon^2 \Omega(t/\epsilon) X(t)=0$ (equation of motion for the zero mode).
\end{itemize}
Then, on each sub-interval, $\mathcal{C}_k$ can be written as the sum of a simplified $\bar{\mathcal{C}}_k$ satisfying the simplified differential equation on this sub-interval, plus a small perturbation $\delta \mathcal{C}_k$. If the exponent $c$ that appears in (\ref{teps}) is appropriately chosen, then, in the singular limit, the $\delta \mathcal{C}_k$ will drop out of the expression for the total $\mathcal{C}(t_1,t_2)$ because they are accompanied by positive powers of $\epsilon$. Then, demanding that the $\epsilon\rightarrow 0$ limit of the complete $\mathcal{C}(t_1,t_2)$ should exist, yields exactly the same condition as the one for the existence of a singular limit of the center-of-mass motion: $Q_{21}(\lambda)=0$. We assume this condition is fulfilled and we obtain in the singular limit:
\begin{align}
\mathcal{C}(t_1,t_2)&=\frac{\sqrt{-\pi t_1 t_2}}{2\sin\alpha\pi}\,\left(Q_{22}(\lambda) J_{a-1/2}(-n t_1)J_{1/2-a}(n t_2)-Q_{11}(\lambda) J_{1/2-a}(-n t_1)J_{a-1/2}(n t_2) \right),\nonumber\\
&\hspace{60mm}\hspace{10mm}t_1<0,\hspace{10mm}t_2>0.\label{exprCtot}
\end{align}

From Wronskian conservation and $Q_{21}(\lambda)=0$ we can write $Q_{11}(\lambda)=q$ and $Q_{22}(\lambda)=-1/q$ (where $q$ can be determined from the center-of-mass motion). We use (\ref{repCfh}) to determine the matching conditions for a basis of solutions \cite{fspw},
\begin{align}
&Y_{1}(t)=\sqrt{-t}J_{a-1/2}(-n t),\hspace{5mm}Y_{2}(t)=\sqrt{-t}J_{1/2-a}(-n t),\hspace{5mm}t<0,\nonumber\\
&Y_{1}(t)=q\sqrt{t}J_{a-1/2}(n t),\hspace{5mm}Y_{2}(t)=-\frac{\sqrt{t}}{q}J_{1/2-a}(n t),\hspace{5mm}t>0.\label{basissol}
\end{align}


\section{The singular limit for the entire string}
\label{total}

Although, for $\lambda >0$, consistent propagation of the string center-of-mass across the singularity guarantees that all excited string modes also propagate in a consistent fashion, small excitations of higher string modes can sum up to yield an infinite total energy \cite{HorowitzSteif} for the whole string. To determine the mode creation we calculate the Bogoliubov coefficients for the string modes related to the transition of the string through the singularity. From (\ref{basissol}) we construct on the one hand a basis of positive and negative frequency modes at early times and on the other hand a basis of positive and negative frequency modes at late times. In light-cone gauge the worldsheet theory inherits the scale invariance of the metric and the Bogoliubov coefficients (that relate the two bases) are independent of $n$:
\begin{equation}
\alpha_n=-\frac{1+q^2}{2 q \,\mathrm{sin}(\alpha\pi)},\hspace{5mm}\beta_n= i\frac{\mathrm{exp}(-i\pi \alpha)+q^2\mathrm{exp}(i\pi \alpha)}{2 q \,\mathrm{sin}(\alpha\pi)}.
\end{equation}
Here, $\alpha=\sqrt{1-4\lambda}/2$. The total mass of the string after crossing the singularity is given by \cite{HorowitzSteif}
\begin{equation}
M=\sum_n n |\beta_n|^2.
\end{equation}
Since the $\beta_n$ are $n$-independent, $M$ can only be finite if $\beta_n=0$ for all $n$. For $\lambda>0$, this cannot be achieved, since $0<\alpha<1/2$ and $q$ is real.


\section{Discussion}
\label{discussion}

Let us first mention that to satisfy background consistency conditions we have chosen to add a dilaton field. The dilaton field in the backgrounds of the type (\ref{genppwave}) takes the form \cite{PRT}
\begin{equation}
\phi=\phi_0+cX^++\frac{d\lambda}2\ln X^+.
\label{dltn}
\end{equation}

For the plane waves that arise as Penrose limits of Friedmann-like cosmologies (positive $\lambda$), it turns out that individual excited string modes propagate consistently across the singularity, whenever the center-of-mass of the string does. In those cases, the dilaton (\ref{dltn}) is actually very large and {\it negative} near the singularity,
and one can expect that free strings are a good approximation as far as propagation across the singularity is concerned (the string coupling is small in the near-singular region). However, for free strings, we find it impossible to maintain a finite total string energy after the singularity crossing, provided that the (scale-invariant) singularity is resolved in a way that does not introduce new dimensionful parameters. The only way out appears to be to allow hidden scales buried at the singular locus (even though the space-time away from the singularity is scale-invariant).

If $\lambda$ is negative, (\ref{dltn}) blows up near $X^+$ (and so does the string coupling) posing a serious threat to the validity of perturbative string theory, and of free string propagation as zeroth order approximation thereto. Our considerations can be seen as a motivation to study these backgrounds in the context of non-perturbative matrix theory descriptions of quantum gravity \cite{quantgrav}.


\section{Acknowledgments}

F.D.R. would like to thank the organizers of the 4-th EU RTN Workshop, in Varna, Bulgaria, for the opportunity to present this work. This research has been supported in part by the Belgian Federal Science Policy Office through the Interuniversity Attraction Pole IAP VI/11, by the European Commission FP6 RTN programme MRTN-CT-2004-005104 and by FWO-Vlaanderen through project G.0428.06.

\end{document}